\documentclass[aps,prb,twocolumn,superscriptaddress,showpacs]{revtex4}
\usepackage{graphicx}
\usepackage{calc}
\usepackage{bm}

\begin{document}

\title{Manifestation of a complex edge excitation structure in the fractional quantum Hall regime at high filling factors}

\author{E.V.~Deviatov}
\email[Corresponding author. E-mail:~]{dev@issp.ac.ru}
 \affiliation{Institute of Solid State
Physics RAS, Chernogolovka, Moscow District 142432, Russia}

\author{A.~Lorke}
\affiliation{Laboratorium f\"ur Festk\"orperphysik, Universit\"at
Duisburg-Essen, Lotharstr. 1, D-47048 Duisburg, Germany}

\author{W.~Wegscheider}
\affiliation{Institut f\"ur Angewandte und Experimentelle Physik,
Universitat Regensburg, 93040 Regensburg,
 Germany}

\date{\today}

\begin{abstract}
We  experimentally study  equilibration across  the sample edge at high fractional filling factors 4/3, 5/3 under experimental conditions, which allow us to obtain  high imbalance conditions. We find a lack of the full equilibration across the edge even in the flat-band situation, where no potential barrier survives  at the sample edge. We interpret this result as the manifestation of complicated edge excitation structure at high fractional filling factors 4/3, 5/3. Also, a mobility gap in the $\nu_c=1$ incompressible strip is determined in normal and tilted magnetic fields.
\end{abstract}

\pacs{73.40.Qv  71.30.+h}

\maketitle

\section{Introduction}

In the fractional quantum Hall effect  (FQHE) regime, an energy gap is present at the Fermi level, separating the many-body ground state and the first excited state. Elementary excitations are supposed to be quasi-electrons or quasi-holes with fractional charge~\cite{laughlin}. The situation becomes more complicated for the systems of finite size. Collective gapless  excitation modes are predicted to exist at the FQHE edge~\cite{wen}. The structure of the edge modes follows the structure of the ground state: there should be several excitation branches for the bulk filling factors $\nu$ that are not of the principal Laughlin sequence~\cite{macdonald,wen}. 

The experimental situation is even more complicated: it was shown for integer~\cite{shklovsky} and fractional~\cite{beenakker,kouwen,cunning} quantum Hall effect regimes, that there is a structure of strips of compressible and incompressible electron liquid at the sample edge.  For the FQHE regime, the strips structure emerges~\cite{chamon,chklovsky} if the edge potential variation is wider than 5-6 magnetic lengths, i.e.,  for most of real edge potentials. Every incompressible strip can be characterized by a local filling factor $\nu_c<\nu$, which corresponds to the integer or fractional quantum Hall state.  The number of strips with integer $\nu_c$ equals to the number of the filled energy levels in the bulk. The number of the strips with fractional $\nu_c$ is determined by the FQHE hierarchical structure~\cite{haldane}, magnetic field and the sample quality. 
Gapless collective modes are predicted~\cite{chamon} at the edges of the every incompressible strip with fractional local filling factor. Physically, they can be understood as oscillations of the strip's shape across the sample edge. Like in the simplest case of a sharp edge profile, there should be several excitation branches for strips with $\nu_c$ that are not of the principal Laughlin sequence.

It is well known that electron transport  across the edge is the most appropriate tool for the investigation of the edge excitations structure~\cite{chang,grayson,fracdens}. However, for narrow fractional incompressible strips, it gives rise the simultaneous excitation of edge modes from different strips~\cite{fracdens}. The influence of several excitation branches at some fractional $\nu_c$ can hardly be separated in this case. 
To avoid this difficuty, samples can be used that have only  integer incompressible strips at the edge and FQHE state in the bulk. 
The easiest way to realize the proposed situation is to use samples of moderate mobility at bulk filling factors $\nu=5/3,4/3$. Only $\nu_c=1 < \nu$ integer incompressible strip can exist at the edge in this case. The potential jump in the integer incompressible strip can be completely eliminated by applying high potential imbalance across it in suitably designed experimental geometry~\cite{alida}.
Thus, transport at higher imbalances could be expected to be sensitive only to the collective edge excitation structure of the bulk fractional state. 
The edge excitation structure  for  $\nu=5/3,4/3$ should consist of several excitation branches~\cite{macdonald}.

Here, we experimentally study  equilibration across the integer incompressible strip with local filling factor $\nu_c=1$ at the sample edge at high imbalances. The bulk is in the quantum Hall state with integer ($\nu=2,3$) or high fractional ($\nu=5/3,4/3$) filling factors. For the fractional ones we find a lack of the full equilibration across the edge even in the flat-band situation, where no potential barrier survives  at the sample edge. We interpret this result as the manifestation of complicated edge excitation structure at high fractional filling factors. Also, a mobility gap in $\nu_c=1$ quantum Hall state is determined in  normal and tilted magnetic fields.

\section{Samples and technique}

The samples are fabricated from  two GaAs/AlGaAs heterostructures with different carrier concentrations and mobilities, grown by molecular beam
epitaxy. One of them (A) contains a 2DEG located 210~nm
below the surface. The mobility at 4K is 1.93 $\cdot
10^{6}$cm$^{2}$/Vs and the carrier density 1.61 $\cdot
10^{11}$cm$^{-2}$, as was obtained from standart magnetoresistance
measurements.
For heterostructure B the  corresponding parameters are 70~nm,
800 000 cm$^{2}$/Vs and 3.7 $\cdot 10^{11}  $cm$^{-2}$. Because of its high carrier concentration, wafer B is used to compare results for integer filling factors at much higher magnetic fields. FQHE states are not achievable for this wafer.

\begin{figure}
\includegraphics*[width=0.9\columnwidth]{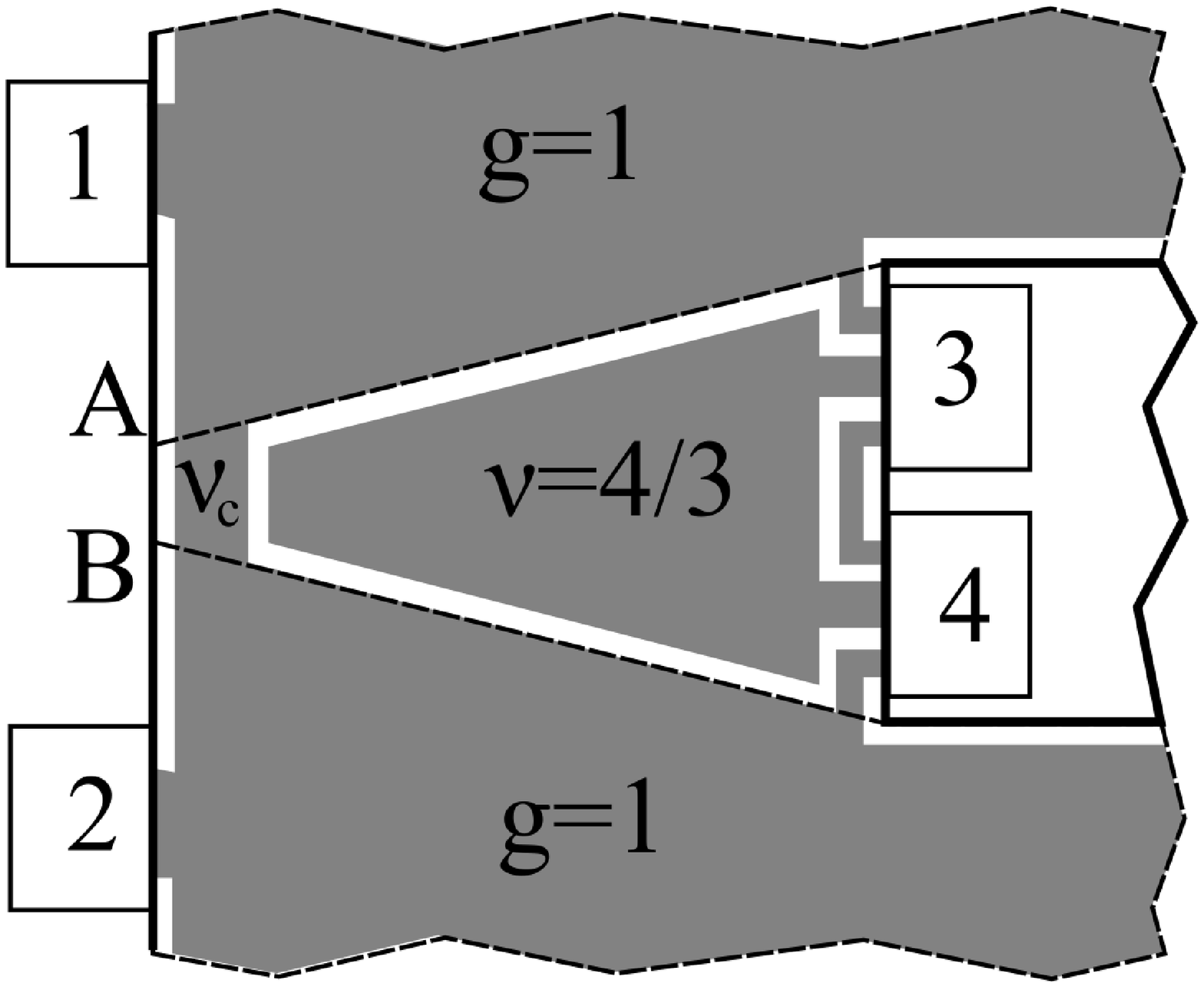}%
\caption{  Schematic diagram of the active sample area. The etched
mesa edges are shown by solid lines, the dashed lines
represent the split-gate edges. The gate-gap region at the outer mesa edge is denoted as AB. Light gray areas are the incompressible
regions at filling factors $\nu$ (in the bulk) and $\nu_c=g<\nu$ (under
the gate and the incompressible strip at the mesa edge).
Compressible regions (white) are at the electrochemical potentials
of the corresponding ohmic contacts, denoted by bars with numbers.
\label{sample}}
\end{figure}

\begin{figure}
\includegraphics*[width=0.9\columnwidth]{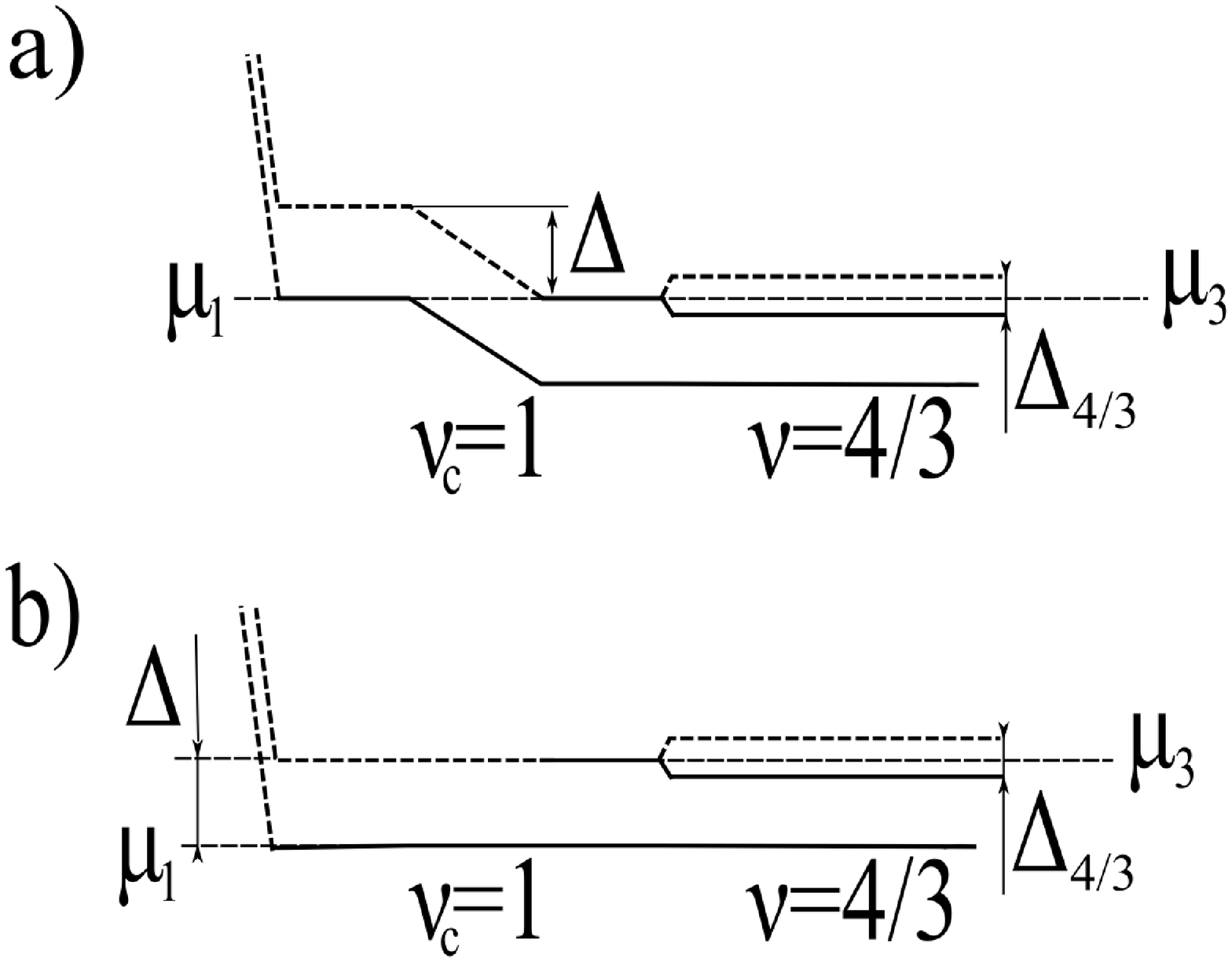}%
\caption{  Schematic diagram of the energy levels in the active sample area. Solid lines represent the filled states, dashed lines are for empty ones. $\Delta$ is the potential jump in the $\nu_c=1$ incompressible strip, $\Delta_{4/3}$ is the energy gap in the bulk. Pinning of the Landau sublevels to the Fermi level (shot-dash) is shown in the compressible regions at electrochemical potentials $\mu_1$ and $\mu_3$. (a) Equilibrium situation $\mu_3=\mu_1$ if no voltage $V$ applied to ohmic contacts 1 and 3. (b) Flat-band conditions for $-eV=-eV_{th}=\mu_1-\mu_3=-\Delta$ ($V>0$, $e$ is the absolute value of the electron charge) 
\label{band}}
\end{figure}

The samples are patterned in the quasi-Corbino sample
geometry~\cite{alida}. Each sample has an etched region
inside, providing a topologically independent inner  mesa edge (Corbino toipology). A split-gate is used to connect these two edges in a controllable way (see Fig.~\ref{sample}). At bulk filling 
factor $\nu$ a structure of compressible and incompressible
strips is present at every edge, see Fig.~\ref{sample}. 
An incompressible quantum Hall state under the gate (filling factor $g<\nu$) is chosen to contact one of the incompressible strips at the outer mesa edge (with local filling factor $\nu_c=g$). In these conditions, some of the compressible
strips (white in Fig.~\ref{sample}) are redirected from inner to outer mesa edge along the split-gate.  Compressible strips  are originally at the electrochemical potentials of the corresponding ohmic contacts.   The gap in the split-gate at
the outer edge (the gate-gap region, denoted as AB in
the figure) has no ohmic contacts inside and is much narrower ($L_{AB}=5 \mu$m in the present experiment) than at the inner one. Both the  macroscopic size of the ungated region at the inner mesa edge  ($\sim 1$mm) and the ohmic contacts  and the ohmic contacts inside it provide a full equilibration among the compressible strips here.    As a result, applying a
voltage between ohmic contacts at outer and inner edges leads to
the electrochemical potential imbalance across the incompressible
strip at local filling factor $\nu_c=g$ at the outer edge, see
Figs.~\ref{sample},\ref{band}.

Magnetocapacitance measurements were performed to characterize the electron system under the gate. By lowering the gate voltage, different quantum Hall states are arising under the gate. The decrease of the electron concentration when approaching the sample edge is a similar process for a soft edge potential. Thus, in magnetocapacitance we obtain not only gate voltage values, corresponding to integer or fractional filling factors $g$ under the gate, but also verify the structure of incompressible strips at the sample edge.  We can confirm in this way, that there is only $\nu_c=1$ incompressible strip for bulk filling factors $\nu=2, 5/3, 4/3$ and there are two integer $\nu_c=1,2$ strips for $\nu=3$. 

In the present paper, the potential imbalance is applied to the outer contacts with respect to the inner ones, i.e. across the incompressible strip at local filling factor $\nu_c=g=1$ in the gate-gap, see Fig.~\ref{band}. The transport takes place across the edge between two spin-split sublevels of the lowest Landau level. They are forming  the bulk incompressible state, which is integer (for $\nu=3,2$) or fractional (for $\nu=5/3,4/3$). The latter situation is demonstrated in Fig.~\ref{band} (a). The potential jump at $\nu_c=1$ is decreasing~\cite{alida} for positive bias, allowing the flat-band situation at $eV=eV_{th}=\Delta$ (see Fig.~\ref{band} (b)) with easy electron transport across the incompressible strip. For $V>V_{th}$, full equilibration is expected~\cite{alida}, leading to a linear $I-V$-characteristic. The corresponding resistance slope $R_{eq}$ can easily be calculated~\cite{beenakker,buttiker}. At negative voltages the potential profile at $\nu_c=1$ is distorted, giving rise to the complicated tunnel branch of the $I-V$ curve.

We study $I-V$ curves in 4-point configuration, by applying \textit{dc} current between a pair of inner and outer contacts and measuring \textit{dc} voltage between another pair of inner and
outer contacts. For the particular contact configuration, equilibrium $I-V$ slope is determined by the relation~\cite{alida}
\begin{equation}
R_{eq}=\frac{h}{e^2}\frac{\nu}{g(\nu-g)}. \label{Req}
\end{equation}
Four-point configuration is used to remove contact resistances (below 100~Ohms in the present samples) from the experimental traces, which allows the most accurate measurements. The linearity of the positive branch of the experimental $I-V$ curves themselves exclude the possibility of Corbino-type or non-linear contacts. Furthermore, the contact behavior is tested separately by  2-point magnetoresistance measurements.  All measurements are performed in a dilution refrigerator with base temperature 30~mK, equipped with a
superconducting solenoid. The results, presented here, are
independent of the cooling cycle.

\section{Experimental results}

\begin{figure}
\includegraphics[width= 0.8\columnwidth]{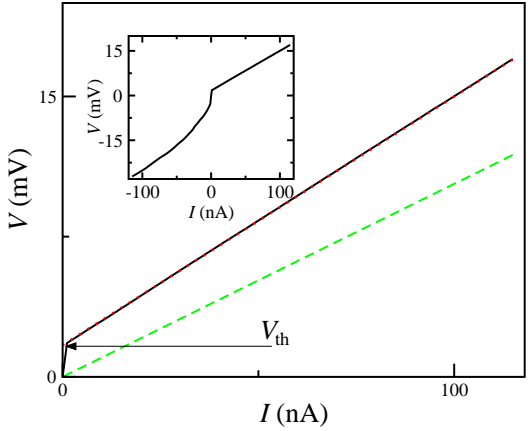}%
\caption{ The positive branch of the $I-V$ curve for the filling factor combination $\nu=4/3, g=1$ (solid curve). The inset also shows the negative branch. The linearity of the positive branch is demonstrated in the main figure. Calculated slope for full equilibration is shown by the dashed line. The threshold voltage $V_{th}$ is determined from the extrapolation to zero current. 
Normal magnetic field equals to 3.58~T. \label{IV43}}
\end{figure}

A typical $I-V$ curve for transport across the integer incompressible strip  is presented in the inset to Fig.~\ref{IV43}. It is strongly non-linear and asymmetric: the positive branch starts from the finite threshold voltage $V_{th}$ and is linear after the threshold; the negative branch continuously goes from zero and is strongly non-linear. This picture is always valid in the framework of the present paper, i.e. for transport across the integer $\nu_c=g$ incompressible strip for both integer and fractional filling factors in the bulk.  The linear behavior of the positive branch above the threshold is demonstrated in a wide voltage range in the main field of Fig.~\ref{IV43}. A linear fit allows to determine the slope $R$ of the positive branch for $V>V_{th}$ with high accuracy. Also, the threshold voltage $V_{th}$ can be obtained from the extrapolation to zero current.

\begin{figure}
\includegraphics[width= 0.75\columnwidth]{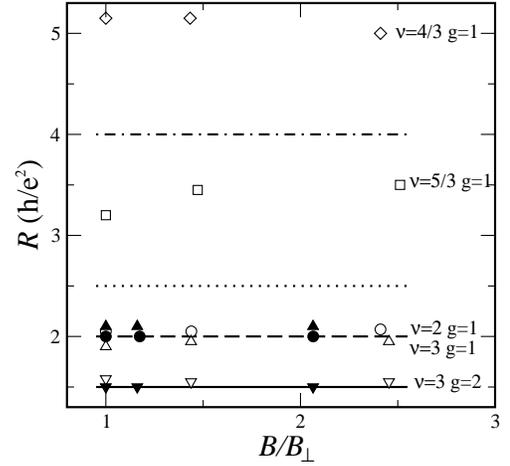}%
\caption{ Slopes of the linear positive branch of the experimental $I-V$ for different filling factor combinations. Filling factor combinations are: $\nu=3, g=2$ (down triangles, $B_\perp=2.38$~T), $\nu=3, g=1$ (up triangles, $B_\perp=2.38$~T), $\nu=2, g=1$ (circles, $B_\perp=3.58$~T), $\nu=5/3, g=1$ (squares, $B_\perp=4.29$~T), $\nu=4/3, g=1$ (diamonds, $B_\perp=5.36$~T). Filled symbols are for the wafer B. Normal magnetic field components are $B_\perp=5.18$~T for $\nu=3$, 7.69~T for $\nu=2$. Lines indicate the calculated slopes for the case of full equilibration: solid line is for $\nu=3, g=2$ and $\nu=3, g=1$, dash is for $\nu=2, g=1$, dotted one is for $\nu=5/3, g=1$, dash-dot is for $\nu=4/3, g=1$.  \label{rdiff}}
\end{figure}

Experimental slopes $R$ are presented in Fig.~\ref{rdiff} for different integer and fractional filling factor combinations. The data are taken at different tilt angles between the magnetic field and the normal to the sample plane. Each dataset is presented as function of the total magnetic field at fixed perpendicular one. Equilibrium slopes, calculated from  Eq.~(\ref{Req}), are denoted by lines. 

The experimental $R$ values are independent of the in-plane magnetic fields for integer filling factor combinations, see Fig.~\ref{rdiff}. They coincide with the calculated lines very well, except for the $\nu=3, g=1$ fillings. The slope of the positive branch for the last one is systematically higher, and is close to one for $\nu=2, g=1$.  This behavior is also supported by the results for the sample B (filled symbols in Fig~\ref{rdiff}), obtained at much higher normal magnetic fields. We can conclude, that the presented behavior is independent of the magnetic field value and is specific to the filling factors only.  The data are temperature independent below 1K.

The experimental situation is more intriguing for the fractional filling factor combinations $\nu=5/3, g=1$ and $\nu=4/3, g=1$. Slopes of the positive  branch are much higher than the calculated values, see Fig~\ref{rdiff}. This fact is also demonstrated directly in the main part of Fig.~\ref{IV43}, there the line with equilibrium slope is depicted. It can also be seen from Fig.~\ref{rdiff}, that there is a weak dependence of $R$ on the in-plane magnetic field. It is different in sign for two bulk fractional filling factors $\nu=4/3$ and $\nu=5/3$ and is more pronounced for the latter one. The data are temperature independent below 0.4~K.  At higher temperatures $R$ values are diminishing with increasing the temperature.

Experimental slopes, exceeding the equilibrium ones, indicate partial equilibration for $V>V_{th}$. This partial equilibration can not be because of higher magnetic fields for fractional fillings, as indicated by good coincidence between the calculation and the experiment for the wafer B at twice higher magnetic fields. Also, different behavior for the filling factor combinations $\nu=3, g=1$ and $\nu=3, g=2$ shows, that the partial equilibration is  only  possible  in the presence of the some structure at the sample edge.

\section{Discussion}

Let us start the discussion from the simplest situation of $\nu=2, g=1$. The only incompressible strip with $\nu_c=1$ is present at the edge in this case and the voltage imbalance is applied across it. Equilibration takes place between two spin-split sublevels of the first Landau level~\cite{relax} (see Fig.~\ref{band} (a), where the edge structure is  similar to one under discussion). It is well known~\cite{mueller,relax} that the equilibration length $l_{eq}$ is determined by both  the potential barrier at local filling factor $\nu_c=1$ and the spin-flip rate. This length  is about 1~mm at low imbalances~\cite{mueller}, that provides high differential resistance ($R\sim l_{eq}/L_{AB}>>R_{eq}$) at $V<V_{th}$. At high imbalances, for $V>0$, the flat-band situation is achieved at $eV=eV_{th}=\Delta_c$, see Fig.~\ref{band} (b). Thus,   there is no potential barrier between compressible strips, allowing electron diffusion along the level. Spin-flip can easily be obtained in this case by photon emission~\cite{relax}. The equilibration length drops below $L_{AB}=5~\mu$m for these reasons~\cite{fracdens}, allowing the full equilibration ($R=R_{eq}$) in our sample. This picture is verified by observation the equilibrium slope of the positive $I-V$ branch for $\nu=2,g=1$, see Fig.~\ref{rdiff}. We can not expect any influence of the in-plane magnetic field on transport in the flat-band situation, at $V>V_{th}$, as  also confirmed in the present experiment.

The situation is more complicated at $\nu=3$. There are two incompressible strips at   local filling factors $\nu_c=1,2$.   Voltage imbalance is applied across one of them, depending on the filling factor $g$.  Potential jump is still survive in the other, that prohibits the flat-band situation across the whole sample edge. There should be only flat band across $\nu_c=g$ incompressible strip at $V>V_{th}$.  For the filling factor combination $\nu=3, g=1$ it is the case:  equilibration takes place between two spin-split sublevels of the first Landau level, like for the $\nu=2, g=1$ fillings, leading to the same experimental slopes (see Fig~\ref{rdiff}) .  For the $\nu=3, g=2$ fillings the equilibration takes place in between two spin-split sublevels of {\em different} Landau levels, across $\nu_c=2$ incompressible strip. This point is crucial for the equilibration: because of the opposite spin component, flowing electrons do not feel the potential jump in the $\nu_c=1$ incompressible strip, producing the full equilibration across the sample edge, see Fig.~\ref{rdiff}.  The partial equilibration, therefore, is a result of some structure of incompressible strips at the edge with $\nu_c\neq g$ and is a sign of such a structure. 

We can not, however, expect any complicated structure of incompressible strips for high fractional filling factors $\nu=5/3$ and $\nu=4/3$. As it is tested from the magnetocapacitance measurements, there is only one incompressible region with $\nu_c=1$ at the edge. This also is in accordance with our estimations from the sample mobility and the magnetic field.  Like in the $\nu=2, g=1$ case, equilibration takes place between two spin-split sublevels of the first Landau level. The bulk state  is the incompressible $\nu=5/3$ or 4/3 FQHE state with gapless collective excitation modes at the edge, see Fig.~\ref{band}. Charge transfer into the edge of the FQHE state is accomplished by their excitation~\cite{wen,chamon}. Thus, partial equilibration in the flat-band conditions (above the threshold) can be achieved only for their selective (partial) excitation. This partial equilibration is independent of the voltage value, leading to the straight line above the threshold. It can be easily understood: in the flat-band conditions electrons are leaving the fractional incompressible state  at the same energy, leading to the same excitation schema at any applied voltage.  For this reason, our experimental results can be regarded as the manifestation of several gapless excitation branches~\cite{chamon,wen} for these high fractional quantum Hall states.  

This conclusion is also supported by the small, but different dependence of the equilibration on the in-plane magnetic field component: the edge excitation structure is different for $\nu=5/3$ and 4/3. In the first case it consists~\cite{chamon,wen} of the integer excitation branch and the quasi-electron fractional one. For $\nu=5/3$ it is represented by the two integer branches and the quasi-hole one. Different spin structure of the integer branches in the latter case can be responsible for the more pronounced in-plane field dependence. The more pronounced temperature dependence of $R$ in comparison with integer fillings also indicates that the effect is strongly connected to the bulk FQHE state.

\begin{figure}
\includegraphics[width= 0.8\columnwidth]{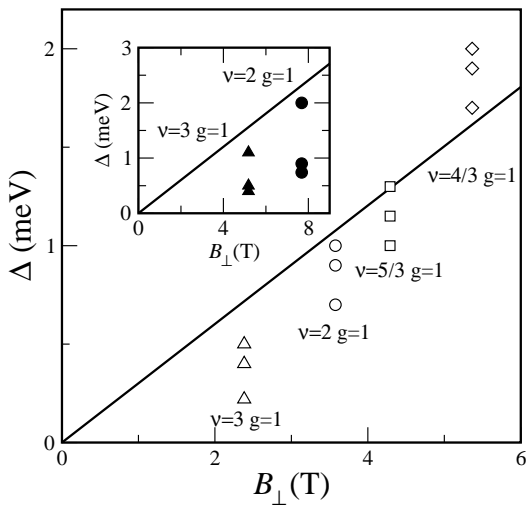}%
\caption{Energy gap in the $\nu_c=1$ incompressible strip as determined from the threshold voltage $V_{th}$ of $I-V$'s for different filling factor combinations. Main figure: $\Delta$ is presented as function of normal magnetic field for the wafer A, for different tilt angles. They are (from down to up): $\theta=0$, $\theta=46^\circ$, $\theta=65.6^\circ$. Solid line indicates the spin gap dependence for the bulk $\nu=1$, measured in Ref.~\protect~\onlinecite{vadik} from the jump of the chemical potential.
Inset: Energy gap in the $\nu_c=1$ incompressible strip, determined for the wafer B. Tilt angles are $\theta=0$, $\theta=31^\circ$, $\theta=61^\circ$. \label{gap}}
\end{figure}

As it can be easily seen, our experiment allows~\cite{alida} to determine the energy gap at local filling factor $\nu_c=1$ by measuring the threshold voltage $V_{th}$ for $g=1$ and different (integer or fractional) $\nu$. Even for the bulk system, energy gap behavior at $\nu=1$ is complicated enough (e.g., compare Refs.~\onlinecite{vadik,shmeller}). Our results, obtained for the incompressible strip at the sample edge, are shown  in Fig.~\ref{gap} as function of the normal magnetic field at different tilt angles. The experimental values are close to the bulk exchange-enhanced gap dependence~\cite{vadik} at $\nu=1$. The normal field dependence is, however, stronger at the edge, see Fig.~\ref{gap}. The gap is rising faster than the linear law, which is very unusual in the bulk. The measured values are even higher than the bulk values at the highest magnetic field. Also, a pronounced in-plane field dependence is present, see Fig.~\ref{gap}.

Before analyzing these results, it is important to understand the nature of the measured gap. We detect the situation $eV_{th}=\Delta$ as the starting point for the current flow. Thus, in the presence of some short-range disorder, it is the mobility gap that is measured in our experiment. It is confirmed by the inset to Fig.~\ref{gap}, where the results are presented for the low-mobility wafer B.  The measured gap values are smaller than for the wafer A, despite the twice higher normal magnetic field values. It indicates the influence of the disorder on the gap measurements in this technique. 

Corresponding measurements for the cyclotron gap at the edge (Ref.~\onlinecite{alida} as well as the present experiment at $\nu=3, g=2$) always show the cyclotron gap values, approximately 20\% smaller than the bulk ones, but with the correct normal and tilted field dependence. Also, the mobility edge deformation can hardly produce mobility gap values, which are higher than the spectrum ones, as we can see for the high-quality wafer A. Thus, the strong  field dependence at $\nu_c=1$ and high gap values can not be ascribed to the change of the mobility edge only. The gap itself shoild rise faster at the edge. 

We can try to attribute the in-plane field dependence it to Zeeman splitting (like it was done in Refs.~\onlinecite{shmeller} for the bulk): 
\begin{equation}
\Delta=\Delta_0(B_\perp)+g^*\mu_BB_{tot}
\end{equation}
where $\Delta_0(B_\perp)$ is the contribution from many-body effects, depends only upon the normal magnetic field $B_\perp$ and the last term is the Zeeman splitting. From our experimental results for the high-quality wafer A, effective g-factor $g^*$ varies from  1.4 (in the lowest normal field) to 0.7 (in the highest one). Assuming a bare g-factor $|g|=0.44$ in GaAs, we can conclude that the number of spin flips varies from 3 to 1. Thus, the number of spin-flips can be higher than one (which is usually attributed to skirmion formation) and depends on the normal magnetic field at the sample edge.  The field dependence of the spin-flip number was not demonstrated for bulk two-dimensional electron systems~\cite{shmeller}. 

The above differencies from the bulk systems  could be because of the different symmetry of the $\nu_c=1$ incompressible strip at the sample edge and the bulk at $\nu=1$. However, there is still an open question about the possibility to obtain the number of the spin-flips from the mobility gap.  The high in-plane field dependence for the wafer B indicates, that the mobility edge change can be crucial in this case.

\section{Conclusion}

In summary, we experimentally study  equilibration across the integer incompressible strip with local filling factor $\nu_c=1$ at the sample edge at high imbalances. The bulk is in the quantum Hall state with integer ($\nu=2,3$) or high fractional ($\nu=5/3,4/3$) filling factors. For the fractional ones we find a lack of the full equilibration across the edge even in the flat-band situation, where no potential barrier survives  at the sample edge. We interpret this result as the manifestation of complicated edge excitation structure at high fractional filling factors. Also, a mobility   gap in $\nu_c=1$ quantum Hall state is determined in it's dependence on normal and tilted magnetic fields.

\acknowledgments

We wish to thank  V.T.~Dolgopolov for fruitful discussions and
A.A.~Shashkin for help during the experiment. We gratefully
acknowledge financial support by the RFBR, RAS, the Programme "The
State Support of Leading Scientific Schools". E.V.D. is also supported by MK-4232.2006.2 and
Russian Science Support Foundation.

\end{document}